# Quantum interface between frequency-uncorrelated down-converted entanglement and atomic-ensemble quantum memory

**Xian-Min Jin[1][†], Jian Yang[1][†], Han Zhang[1], Han-Ning Dai[1], Sheng-Jun Yang[1], Tian-Ming Zhao[1], Jun Rui[1], Yu He[1], Xiao Jiang[1], Fan Yang[2], Ge-Sheng Pan[1], Zhen-Sheng Yuan[1,2], Youjin Deng[1], Zeng-Bing Chen[1], Xiao-Hui Bao[1,2], Bo Zhao[3], Shuai Chen[1] and Jian-Wei Pan[1,2]**

[1]Hefei National Laboratory for Physical Sciences at Microscale and Department of Modern Physics, University of Science and Technology of China, Hefei, Anhui, 230026, PR China

[2]Physikalisches Institut, Universität Heidelberg, Philosophenweg 12, D-69120 Heidelberg, Germany

[3]Institute for Theoretical physics, University of Innsbruck, A-6020 Innsbruck, Austria

[†]These authors contributed equally to this work.

**Photonic entanglement source and quantum memory are two basic building blocks of linear-optical quantum computation[1,2] and long-distance quantum communication[3,4]. In the past decades, intensive researches have been carried out, and remarkable progress, particularly based on the spontaneous parametric down-converted (SPDC) entanglement source[5,6,7] and atomic ensembles[8,9,10], has been achieved. Currently, an important task towards scalable quantum information processing (QIP) is to**

efficiently write and read entanglement generated from a SPDC source into and out of an atomic quantum memory[11,12]. However, the physical realization of such a quantum interface remains rather challanging, mainly due to the notorious mismatch between the working bandwidth of the SPDC source and atomic quantum memory. Albeit narrow-band SPDC sources, whose linewidth is approaching to the requirement of atomic memory, have been demonstrated very recently [13,14,15], the exact frequency correlation between the entangled photons makes them neither suitable for storage nor practically useful for scalable QIP[16,17]. Here we report the first experimental realization of a quantum interface by building a 5 MHz frequency-uncorrelated SPDC source and reversibly mapping the generated entangled photons into and out of a remote optically thick cold atomic memory using electromagnetically induced transparency. The frequency correlation between the entangled photons is almost fully eliminated with a suitable pump pulse. The storage of a triggered single photon with arbitrary polarization is shown to reach an average fidelity of 92% for 200 ns storage time. Moreover, polarization-entangled photon pairs are prepared, and one of photons is stored in the atomic memory while the other keeps flying. The CHSH Bell's inequality[18] is measured and violation is clearly observed for storage time up to 1 μs. This demonstrates the entanglement is stored and survives during the storage. Our work establishes a crucial element to implement scalable all-optical QIP, and thus presents a substantial progress in quantum information science.

Photonic entanglement source and quantum memory are the basic ingredients of scalable all-optical QIP. Currently, the most widely used photonic entanglement source is the spontaneous parametric down converted (SPDC) entanglement source, and the most promising quantum memory is based on atomic ensembles using electromagnetically induced transparency (EIT). In last decade, significant progress has been achieved in these two fields. Up to six-photon entanglement[19,20,21,22] and ten-qubit entanglement[23] have already been demonstrated. Storage of single photons[24,25,26] and non-classical light[27,28] has been realized with the coherence time extended to millisecond regime[29,30].

To implement all-optical QIP in a scalable way, entangled photons should be stored in quantum memory and then be retrieved out asynchronously, since otherwise the required resources will increase exponentially with the computation scales and the communication distance. However, a direct storage of the SPDC entangled photons in the atomic memory is extremely challenging because of the notorious broadband nature of the SPDC entanglement source (THz), far beyond the working bandwidth of the atomic quantum memory (MHz). To remedy this problem, significant efforts have been made towards developing narrow-band SPDC entanglement sources[31,32,33], and a bandwidth of MHz has been achieved very recently[13,14,15]. However, these entangled photons are randomly distributed in a continuous beam and cannot be efficiently stored in atomic ensembles. Further, they are exactly frequency correlated, leading to a very low efficiency of multi-photon interference[16,17]. As a consequence, these narrow-band SPDC sources are of no practical interest yet for scalable QIP[2,11].

Here we report an experimental realization of a quantum interface between a narrowband frequency-uncorrelated SPDC entanglement source and an atomic quantum memory, which enables efficient mapping of frequency-uncorrelated entangled photons into and out of an atomic quantum memory using EIT. To such a challenging goal, four major improvements are achieved in our experiment: (i) by inserting a compensation crystal in the cavity to fulfill the double resonance for orthogonally polarized photons, we obtain a bright frequency-tunable narrow-band source; (ii) by choosing a proper pulse regime and controlling the pulse duration of the pump light, we eliminate the frequency correlation between the entangled photons; (iii) by employing a rectangular dark magneto optical trap (MOT), we improve the optical depth of the atomic ensemble up to 55, and thus can gain a high retrieval efficiency and a full compression of the entangled-photon wave packets into the atomic ensemble with negligible leakage; (iv) by designing an interferometer configuration with two EIT processes to coherently store the horizontal (H) and vertical (V) polarization modes, we successfully store entanglement in an arbitrary unknown state.

**Preparation of the entanglement source and atomic memory**

Our experimental setup is illustrated in Fig. 1, where entangled photons are generated in the source lab and stored in the memory lab located 10 m away. In the source lab, an ultraviolent (UV) pump light is applied to a periodically poled KTiOPO4 (PPKTP) crystal inside a cavity with linewidth $\gamma$ about 5 MHz. This produces narrow-band entangled photon pairs with the central frequency $\omega_0$ tuned to match the D1 line of $^{87}Rb$.

Single-mode output is realized by employing an actively stabilized filter cavity (See Methods). The cavity-enhanced twin photons are subject to a polarization beam splitter (PBS) and interfere with each other, resulting in an entangled state described by

$$|\Psi\rangle_{1,2} \propto \int d\omega_1 d\omega_2 \psi(\omega_1,\omega_2)(a_{1H}^\dagger(\omega_1)a_{2V}^\dagger(\omega_2) + a_{1V}^\dagger(\omega_1)a_{2H}^\dagger(\omega_2)) \quad (1),$$

where $\psi(\omega_1,\omega_2)$ is the frequency function and $a_{1,2;H,V}^\dagger(\omega)$ is the creation operator for photon 1 or 2 with H or V polarization. Photon 1 is flying to a polarization analyzer through a 60 m single-mode fiber, while photon 2 is directed to the memory lab through 20 m single-mode fiber for further operation. In the following we shall call photon 1 the flying photon and photon 2 the signal photon, respectively.

In the memory lab, a cigar-shaped cold atomic cloud of $\sim 10^8$ $^{87}$Rb atoms is confined in a rectangular MOT and serves as the quantum memory for the signal photon. Dark MOT technique[34] is applied to further increase the optical density to up to 143. In order to reduce the collisions between the cold atoms and the thermal background particles, we choose to collect experimental data with optical depth 55 in a vacuum of $5\times10^{-9}$ mbar. Two spatial modes U and D passing through the long-axis direction of the atomic cloud are employed as two separated atomic ensembles[26,35,36]. The standard EIT experiment is performed, where a probe light is applied in the U/V mode and a control light is applied at an angle of 2 degree (see Fig. 2b). For a control light with a Rabi frequency of 12.6 MHz, we observe an EIT window of 5.5 MHz and a delay time of 200 ns, which corresponds to a delay-bandwidth product of 7.

In order to demonstrate the storage capability of the atomic memory, we cut the UV

pump light into pulses with duration $T_p = 100$ ns. The resulted signal photon pulse has duration about 200 ns (full-width at half-maximum about 100 ns) and a bandwidth of 5 MHz, which well match the delay time and EIT window of the atomic memory. We then perform the stop light experiment on the V polarization mode of this pulse by adiabatically switching off the control light after the pulse entered ensemble 1 (See Methods). Because of the high optical depth and delay-bandwidth product, we can fully compress such a pulse into the atomic ensemble with negligible leakage. After a controllable delay, the stored pulse is read out and detected, where the measured leakage efficiency and the overall storage efficiency for different storage time are shown in Fig. 2c.

**Preparation of frequency-uncorrelated entangled photons**

In general, the entangled photons created from the SPDC source are frequency correlated, which means that the photon pulse in one mode is not Fourier-transform-limited and thus unsuitable for storage[8,12]. Even worse, the frequency correlation will significantly lower the visibility of multi-photon interference for photons emitted from independent sources, and consequently decrease the fidelity of the generated multi-photon entanglement[16,17]. Therefore, eliminating the frequency correlation between the entangled photons is of key importance to scalable QIP[2,11].

For narrow-band entanglement sources, due to the extremely narrow band nature of the cavity, the frequency correlation can be eliminated by choosing a suitable pumping pulse with a linewidth larger than that of the cavity, as explained following. The two-photon state in the frequency domain can be written as

$$\psi(\omega_1,\omega_2) \propto \frac{1}{(\omega_1-\omega_0)+i\frac{\gamma}{2}} \cdot \frac{1}{(\omega_2-\omega_0)+i\frac{\gamma}{2}} \cdot \phi(\omega_1+\omega_2-2\omega_0) \tag{2}$$

where $1/(\omega_{1,2}-\omega_0+i\gamma/2)$ is the spectrum response function of the cavity, and $\phi(\omega)$ is the frequency spectrum of the pump light. The spectrum response function is centered at $\omega_0$ and decreases very rapidly away from $\omega_0$. Therefore, when the bandwidth of $\phi(\omega)$ is much larger than the linewidth $\gamma$, the frequency of the two photons will be mainly determined by the spectrum response function, and the frequency correlation will be eliminated. The frequency correlation of two entangled photons can be quantified by a number associated with the visibility of multi-photon interference[18] $V=\xi/\kappa$, with

$$\xi = \int d\omega_1 d\omega_2 d\omega_3 d\omega_4 \psi(\omega_1,\omega_2)\psi(\omega_3,\omega_4)\psi^*(\omega_1,\omega_4)\psi^*(\omega_3,\omega_2)$$

and $\kappa = \int d\omega_1 d\omega_2 d\omega_3 d\omega_4 |\psi(\omega_1,\omega_2)\psi(\omega_3,\omega_4)|^2$. A straightforward calculation yields $V=1$ for two frequency-uncorrelated photons. For a Gaussian pumping light with bandwidth $\sigma$, i.e., $\phi(\omega)=(1/\sqrt{2\pi}\sigma)\exp(-\omega^2/2\sigma^2)$, the numerical calculation gives $V=97\%$ for $\sigma=12.5$ MHz and 80% for 3.7 MHz, which corresponds to a pump pulse of $T_p=30$ ns and $T_p=100$ ns, respectively.

The frequency correlation can be observed by measuring the two-photon distribution in time domain, obtained from Fourier transformation of Eq. (2). The two-photon state in time domain is $\psi(t_1,t_2) \propto \exp(-\gamma|t_1-t_2|/2)$ in the continuous pump limit $\phi(\omega)=\delta(\omega)$, where frequency is exactly correlated, while it is $\psi(t_1,t_2) \propto \exp(-\gamma t_1/2)\exp(-\gamma t_2/2)$ in the short pump limit $\phi(\omega)=const$ without frequency correlation[13]. Thus, the frequency correlation can be readily observed by plotting two-photon distribution $P(t_1,t_2) = |\psi(t_1,t_2)|^2$ in the $(t_1,t_2)$ plane, where the frequency correlation is reflected by a peak along the diagonal line (see Fig. 3a and 3b).

In the experiment, we choose $T_p = 30$ ns and $T_p = 100$ ns, and measure the two-photon distribution by detecting the flying photon and the signal photon in H and V basis, respectively. The theoretical calculations and experimental results agree with each other, as shown in Fig. 3c-3f. We also measure two-photon distribution after storage of the signal photon (see details below). The results are shown in Fig. 3g and 3h. It can be readily seen that for $T_p = 30$ ns, the frequency uncorrelated feature is well preserved. While for $T_p = 100$ ns, the peak along the diagonal line disappears, which implies that the frequencies of the two photons become less correlated after storage. This may be due to the narrow EIT transparency window which also acts as a frequency filter.

The following experiments for storage purpose are carried out for $T_p = 30$ ns with the frequency correlation almost fully eliminated. In this case, the entangled photons are purely entangled in polarization space and can be expressed as $|\Psi\rangle_{1,2} = 1/\sqrt{2}(|H\rangle_1|V\rangle_2 + |V\rangle_1|H\rangle_2)$.

**Storage a single photon with arbitrary polarization**

We first store a polarized signal photon triggered by the flying photon, and the polarization is predetermined to H by a polarizer before PBS1. The signal photon is stored in ensemble 2 by adiabatically switching off the control light and is then read out by increasing its strength after a controllable delay. Note that, for $T_p = 30$ ns, the full length of the signal photon is about 200 ns with a linewidth of 5 MHz due to the long exponential decay tail, and is thus suitable for storage.

To demonstrate that the single-photon nature is preserved in storage, we measure the cross correlation between the two photons, expressed as $g_{13}^{(2)} = p_{13}/(p_1 p_3)$ with $p_1$ ($p_3$) the probability of detecting a photon on detector D1 (D3) and $p_{13}$ the coincident probability between D1 and D3. The single-photon quality of the signal photon can be estimated by $\alpha = 4/(g_{13}^{(2)} - 1)$, which is zero for pure single photon and 1 for weak coherent state[29,35]. The measured $g_{13}^{(2)}$ as a function of storage time is shown in Fig 4a. The storage time for the single photon can be estimated to be 2 μs, when the cross correlation drops below 5.

To demonstrate that a single photon with arbitrary polarization state can be well stored, we change the polarization of the signal photon by tuning the angle of the polarizer, and prepare the following six states as the initial states to be stored (See Methods). They are the linear polarization states $|H\rangle$, $|V\rangle$, $|+\rangle = (|H\rangle + |V\rangle)/\sqrt{2}$, $|-\rangle = (|H\rangle - |V\rangle)/\sqrt{2}$ and the circular polarization states $|R\rangle = (|H\rangle + i|V\rangle)/\sqrt{2}$, $|L\rangle = (|H\rangle - i|V\rangle)/\sqrt{2}$.

In order to coherently store arbitrary-polarization states, the H and V modes of the signal photon are coherently separated by PBS1, and respectively directed to the two spatial modes U and D, of which each ensemble serves as a quantum memory for a polarization mode. The two spatial modes share the same control light. This arrangement is superior not only for simplicity but also for the fact that phase noises in the two EIT processes induced by phase drift of two control fields will be suppressed. Note that the coherence time of the retrieved photon is tunable and has a width of tens of nanosecond, thus it is very easy to achieve good temporal overlap for the two spatial modes at PBS2.

In the experiment, the photonic state $|\psi_{in}\rangle = \alpha|H\rangle + \beta|V\rangle$ is mapped into the atomic ensemble as a superposition of atomic state $|\psi\rangle_a = \alpha|H\rangle_a + \beta|V\rangle_a$, where $|H\rangle_a = |0\rangle_U |1\rangle_D$, $|V\rangle_a = |1\rangle_U |0\rangle_D$ with $|0\rangle_{U,D}$ the vacuum state and $|1\rangle_{U,D}$ the single-excitation collective excited state in ensemble U or D. The stored state is read out after a controllable delay and in ideal case the retrieved state $|\psi\rangle_{out}$ is the same as the input.

The performance of the quantum memory of the photonic qubit can be estimated by the storage fidelity $F = |\langle \psi_{in} | \psi_{out} \rangle|^2$, as measured with a state analyzer which projects the retrieved state $|\psi_{out}\rangle$ to $|\psi_{in}\rangle$ and its orthogonal state $|\psi_{in}^\perp\rangle$. The storage fidelity after 200 ns storage time is shown in Fig. 4b-4d, where the fidelity varies from (88.1±3.5)% to (98.9±1.0)% for different initial states. We obtain an average fidelity of (92.4±1.2)% by averaging over the six initial states. This demonstrates the high quality of the quantum memory.

**Storage of entanglement**

To demonstrate the storage of entanglement, we take away the polarizer before PBS1. In this case, the polarization state of the signal photon is unknown, since it is entangled with the flying photon, which is propagating in the 60 m fiber and will not be detected until the signal has been fully stored in the atomic ensemble. This is distinct from the single-photon storage, where the polarization is already determined before storage. In the experiment, the photonic entangled state $|\Psi_{in}\rangle_{1,2} = (1/\sqrt{2})(|H\rangle_1|V\rangle_2 + |V\rangle_1|H\rangle_2)$ is mapped to an atom-photon entanglement $|\Psi\rangle_{1,a} = (1/\sqrt{2})(|H\rangle_1|V\rangle_a + |V\rangle_1|H\rangle_a)$ during storage, and is then read out again as a photonic entangled state $|\Psi_{out}\rangle_{1,2} = (1/\sqrt{2})(|H\rangle_1|V\rangle_2 + |V\rangle_1|H\rangle_2)$.

To justify the performance of the entanglement storage, we first measure the polarization correlation between the flying photon and the retrieved signal photon. The polarization state of the flying photon is projected onto $|H\rangle$ and $|+\rangle$ basis and the polarization state of the signal photon is projected to $\cos\theta|H\rangle + \sin\theta|V\rangle$, with $\theta$ the projection angle. The normalized coincidence versus the projection angle for a storage time of 200 ns is shown in Fig. 4e., which yields a visibility about 81%. This is well beyond the limit 70.7% to violate Bell's inequality.

To further demonstrate the survival of the polarization entanglement during storage, we measure the violation of Clauser-Horne-Shimony-Holt (CHSH) type Bell's inequality[18]. Correlation function $E(\theta_1, \theta_2)$, where $\theta_1$ and $\theta_1'$ ($\theta_2$ and $\theta_2'$) are the polarization angles of the flying (signal) photon, are measured, and we calculate the quantity of $S = |E(\theta_1, \theta_2) - E(\theta_1, \theta_2') + E(\theta_1', \theta_2) + E(\theta_1', \theta_2')|$. Local hidden variable theories yield $S \leq 2$ for any angle arrangement. We set $(\theta_1, \theta_1', \theta_2, \theta_2') = (0°, 45°, 22.5°, 67.5°)$ and measure $S$ for the following three cases: (i) the two photons are both in the source lab (Local); (ii) the signal photon is stored for 200 ns, and (iii) for 1 μs. The observed $S$ values are shown in Table 1. The CHSH inequality is violated for all the three cases. This proves that the entanglement is stored and survives at least for a storage time of up to 1 μs.

**Outlook**

In summary, we have successfully demonstrated an efficient and reversible mapping of the entanglement generated from SPDC sources into and out of atomic ensembles using EIT. The extension of our work to storage of many entangled photons could be realized by employing more spatial modes or temporal modes[37]. The storage time can be extended to milliseconds by using clock state and suppressing the spin-wave dephasing[29,30]. The methods of preparing frequency-uncorrelated narrow-band entangled pairs and storing them in a quantum memory open up many exciting possibilities for experimental investigations of many new all-optical quantum information protocols based on storage of photonic entanglement, such as preparation of large-scale graph states[38], demonstration of quantum repeater based on storage of heralded entangled pairs[39], implementation of linear-optical one-way quantum computing via storage of a cluster state[40,41,42], implementation of high-precision quantum measurement via storage of photonic NOON state[43]. Our work provides a quantum interface between the SPDC entanglement source and the atomic memory, and thus presents a substantial progress towards scalable QIP.

**Methods**

**Preparation of narrow-band entanglement source.**

The cavity mainly contains three parts in sequence, i.e., a nonlinear crystal, a tuning crystal and an output coupler. The nonlinear crystal is a 25mm-long type-Ⅱ PPKTP crystal, whose operational wavelength is designed to match the D1 transition line of $^{87}Rb$.

The first side of the PPKTP is high-reflection coated (R>99.99% at 795 nm) to form the double-resonant cavity with the concave output coupler (R =97% at 795 nm) of 15-cm curvature. A 5mm-long KTP crystal whose axis is perpendicular to that of the PPKTP crystal functions as a tuning crystal. Consequently, by tuning the temperature of the crystals cooperatively, the operational wavelength can cover almost the whole range of the phase-matching bandwidth of the nonlinear crystal with optimal phase matching, and meanwhile fulfills the double-resonant condition of type-II cavity-enhanced SPDC. The temperature of the crystals is controlled with a temperature fluctuation of about 0.002 K. The cavity is actively stabilized by the Pound-Drever-Hall strategy and the measured finesse is 170. A single piece of fused silica of about 6.35mm is designed as the filter cavity with a finesse of 30 to select the single-mode output. The bandwidth of the single-mode narrow-band source is about 5MHz. The narrow-band entanglement is generated by interfering with the twin photons generated from the cavity-enhanced SPDC source. The fiber-coupled coincident rate is about 50 counts per second per mw pump power and MHz bandwidth, which is nearly 20 times higher than the previous narrow-band source[13]. With a periodically driven acousto-optic modulator (AOM), we cut the pump laser into pulsed form and obtain a pulsed source with an interference visibility of 25. In this way, we prepare bright stable and pulsed narrow-band photonic entanglement for the atomic quantum memory.

**Experimental details of storage.**

In the experiment, the dark MOT is loaded for 20 ms at a repetition rate of 47Hz. The magnetic field, the cooling and the re-pumping lasers are then quickly switched off (within 10 μs). Within another 1.3 ms, 200 experimental trials are performed. In every trial, a pumping laser ($|5S_{1/2}, F=2\rangle \to |5P_{3/2}, F=2\rangle$) is first switched on to prepare the cold atoms in the initial ground state $|5S_{1/2}, F=1\rangle$; Narrowband entanglement source in the source lab is triggered to be pulsed by a user-defined TTL signal from the memory lab; When the signal photon arrives at the memory lab, the control laser is switched on for 400 ns, which is sufficient to compress the whole signal photon pulse into the cold atoms. After a programmable storage time, the control laser is switched on and lasts for 600 ns. Each experimental data point is obtained by averaging over $3.2 \times 10^7$ the experimental trials. In order to minimize the phase noise between the control field and the signal field, we first lock the pump laser (MBR) (master laser) of the source to the transition $|5S_{1/2}, F=1\rangle \to |5P_{1/2}, F=2\rangle$ using the saturated absorption technique. Then, we make a beat measurement with a semiconductor laser (slave laser). A home-built phase lock module compares the obtained beat signal with a standard 6.8 GHz signal and give a fast feedback signal to drive the current of the slave laser. As a consequence, frequency drift between the master and the slave laser is controlled to be within 1 Hz.

**Establish the quantum memory for arbitrary polarization.**

As shown in Fig.1, our quantum memory consists of a built-in Mach-Zehnder interferometer, a cloud of cold $^{87}$Rb atoms trapped by dark MOT, and a strong control field. The signal photon with an arbitrary initial polarization state $|\Psi\rangle = \alpha|H\rangle + \beta|V\rangle$ is coherently split into two spatial mode D (U) with a probability of $|\alpha|^2$ ($|\beta|^2$) and directed into the two atomic ensembles. To meet the complete EIT condition, the polarizations in both spatial modes are transferred into $\sigma^+$, so does the control field. The two spatial modes of the signal photon, as illustrated in Fig.2b, cross each other at cigar-shaped cold atomic ensembles and have an angle of $2°$ with the control field. The speed of the signal photon in each spatial mode is slowed down to $2\times10^4$ m/s in the presence of control field (Rabi frequency 12.6 MHz). After being fully compressed into the atomic ensembles, the signal photon is converted into a pure atomic superposition state $\alpha|0\rangle_U|1\rangle_D + e^{i\varphi_1}\beta|1\rangle_U|0\rangle_D$ by adiabatically switching off the control field, where the phase $\varphi_1$ is the phase difference between the two modes from PBS1 to the two atomic ensembles. After a programmable delay, the signal photon is read out by switching on control field. Additional pairs of HWPs and QWPs rotate $\sigma^+$ polarization back to V and H. The signal-photon state at PBS2 becomes $|\Psi\rangle = \alpha|H\rangle + e^{i(\varphi_1+\varphi_2)}\beta|V\rangle$, with $\varphi_2$ the phase difference between the two modes from the atomic ensembles to PBS2. A phase probe laser (wavelength 780nm) with $|+\rangle$ polarization is coupled into the interferometer along the counter-propagating direction of the signal photon. At the output of PBS1, the polarization at the $|+\rangle/|-\rangle$ basis is analyzed and the results are sent to the feedback control system (see Fig. 1), which yields a feedback signal for Piezo to set $\varphi_1+\varphi_2$ to zero. The instability can be suppressed within λ/28. In our experiment, the main background noise comes from the leakage of the control field ($|5S_{1/2}, F=2\rangle \to |5P_{1/2}, F=2\rangle$), which is 6.8 GHz away from the signal photon ($|5S_{1/2}, F=1\rangle \to |5P_{1/2}, F=2\rangle$). To filter the control

field from the retrieved signal photon, three methods are employed: (i) the angle between the signal field and the control field is an effective spatial filter, which contributes about 140dB; (ii) a Fabry-Perot cavity contributes about 20 dB. The cavity is designed to be compact and the temperature is controlled to be of the precision of about 0.002 K; (ii) A pure $^{87}Rb$ hot cell (length 10 cm) contributes about 30 dB. The cell is heated to 80 °C for the desired optical depth and the atoms are prepared at $|5S_{1/2}, F=2\rangle$, which leads to strong resonance absorption of the control field but transparency to the signal field. As a result, the noises from the control field are suppressed to the same level as the noises of the single-photon detectors.

**Acknowledgements**

This research leading to the results reported here was supported by the Chinese Academy of Sciences, the National Fundamental Research Program of China under Grant No.2006CB921900, and the National Natural Science Foundation of China.

**Competing financial interests**

The authors declare that they have no competing financial interests.

**Figure Captions:**

**Figure 1 Schematic view of the experimental scheme.** The narrow-band SPDC-entanglement source is a periodically poled KTiOPO4 crystal (PPKTP) inside a cavity and located in the source lab. The photonic entangled pair is generated by applying a ultraviolent pump light. The flying photon is coupled into a 60 m single-mode fiber and guided to a polarization state analyzer, which consists of half wave plate (HWP), quarter wave plate (QWP), polarization beam splitter (PBS) and single photon detectors. The signal photon is directed through 20 m fiber to the memory lab at a distance of 10 m. In the memory lab, a cold atomic cloud trapped in MOT serves as the quantum memory for the signal photon, with spatial modes U and D selected as two ensembles for storage of the horizontal (H) and the vertical (V) polarization mode, respectively. The path-length difference between U and D modes is set at zero and actively stabilized. The two polarization modes are both transferred to be $\sigma^+$ for storage. A strong control light is applied at an angle of 2 degree relative to the two spatial modes. The signal photon is stored and read out of the atomic ensemble by changing the strength of the control light.

The retrieved signal photon is transmitted through a Fabry-Perot cavity and atomic filter cell in order to absorb the leakage from the control light, and then subject to a polarization analyzer for state analysis.

**Figure 2 Characterizations of storage medium. a**. Transmission spectra of a coherent 50-μs-long probe light versus the probe detuning from ($|5S_{1/2}, F=1\rangle \to |5P_{1/2}, F=2\rangle$) transition in the presence (red circles) and the absence (blue circles) of the control field. The optical depth (OD) deduced from the measured absorption profiles is 55. An EIT widow of 5.5 MHz is observed for a control light with Rabi frequency of 12.6 MHz. Each data point represents an average of 20 experimental trials. **b.** Illustration of the laser configuration. The inset shows the relevant atomic levels of $^{87}$Rb and the associated light fields. **c.** Experimental distribution shapes as a function of time, showing the storage and the retrieval of the signal photon without subtracting background. We register the coincidence between D1 and D3 before storage, during storage and after retrieval. The leakage efficiency is defined as the ratio of the coincidence during storage and before storage, and the retrieval efficiency is defined by the ratio of the coincidence after retrieval and before storage.

**Figure 3 Normalized density of the two-photon distribution versus the detection time (ns). a, b,** Theoretical results for perfect frequency correlation and no frequency correlation, respectively. **c, d,** Theoretical results for $T_p = 100$ ns and $T_p = 30$ ns. **e, f,** Experimental results without storage for $T_p = 100$ ns and $T_p = 30$ ns. **g, h,** Experimental results after storage for $T_p = 100$ ns and $T_p = 30$ ns.

**Figure 4 Performance of the quantum memory for polarized photons and entanglement. a,** Cross correlation as a function of time. **b,c,d,** Storage fidelities for the six initial polarization states after 200 ns storage time. The storage fidelities are $0.954\pm0.026$ ($|H\rangle$), $0.989\pm0.010$ ($|V\rangle$), $0.909\pm0.027$ ($|+\rangle$), $0.889\pm0.037$ ($|-\rangle$) $0.920\pm0.031$ ($|R\rangle$) and $0.881\pm0.035$ ($|L\rangle$). **e,** Normalized polarization correlation for the retrieved entanglement. The red (blue) curve represents that the polarization of the flying photon is fixed to $|H\rangle$ ($|+\rangle$). Statistical error bars are displayed.

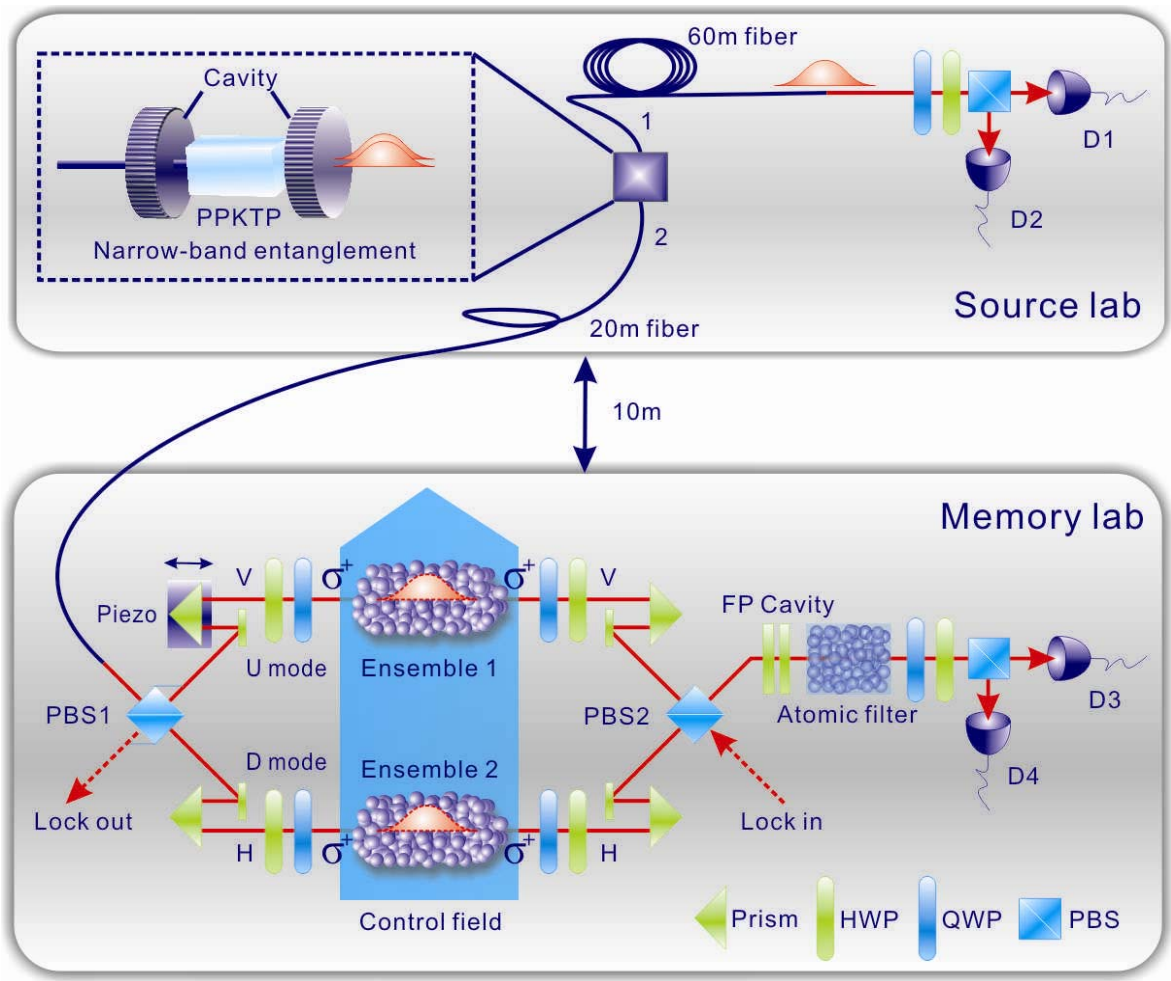

Figure-1

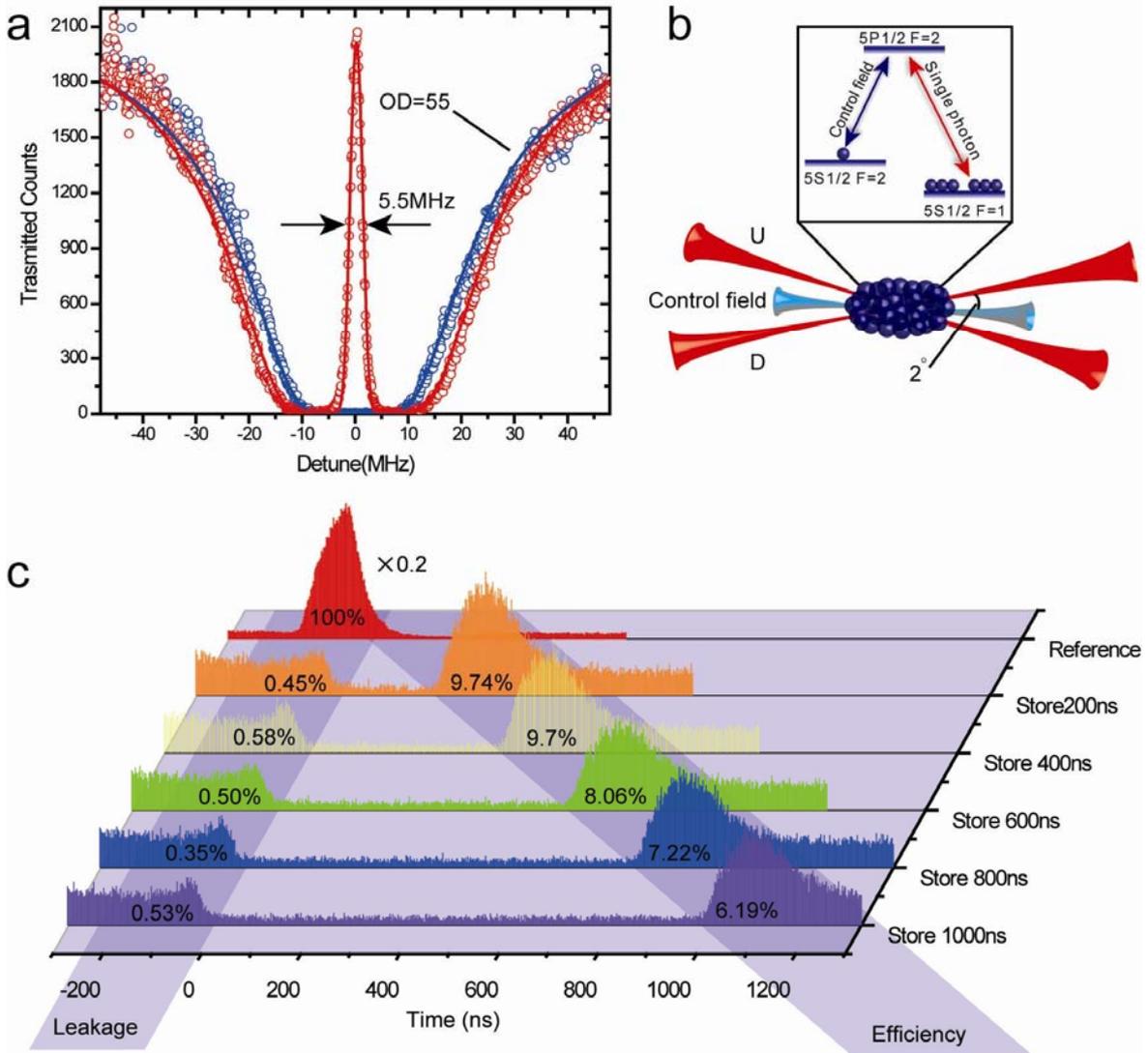

Figure-2

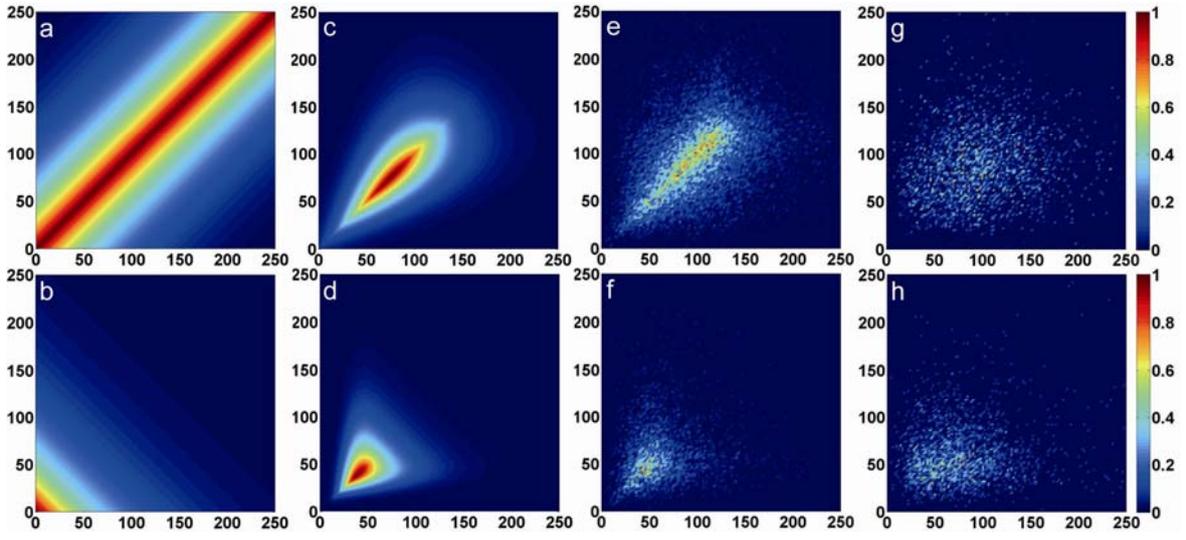

Figure-3

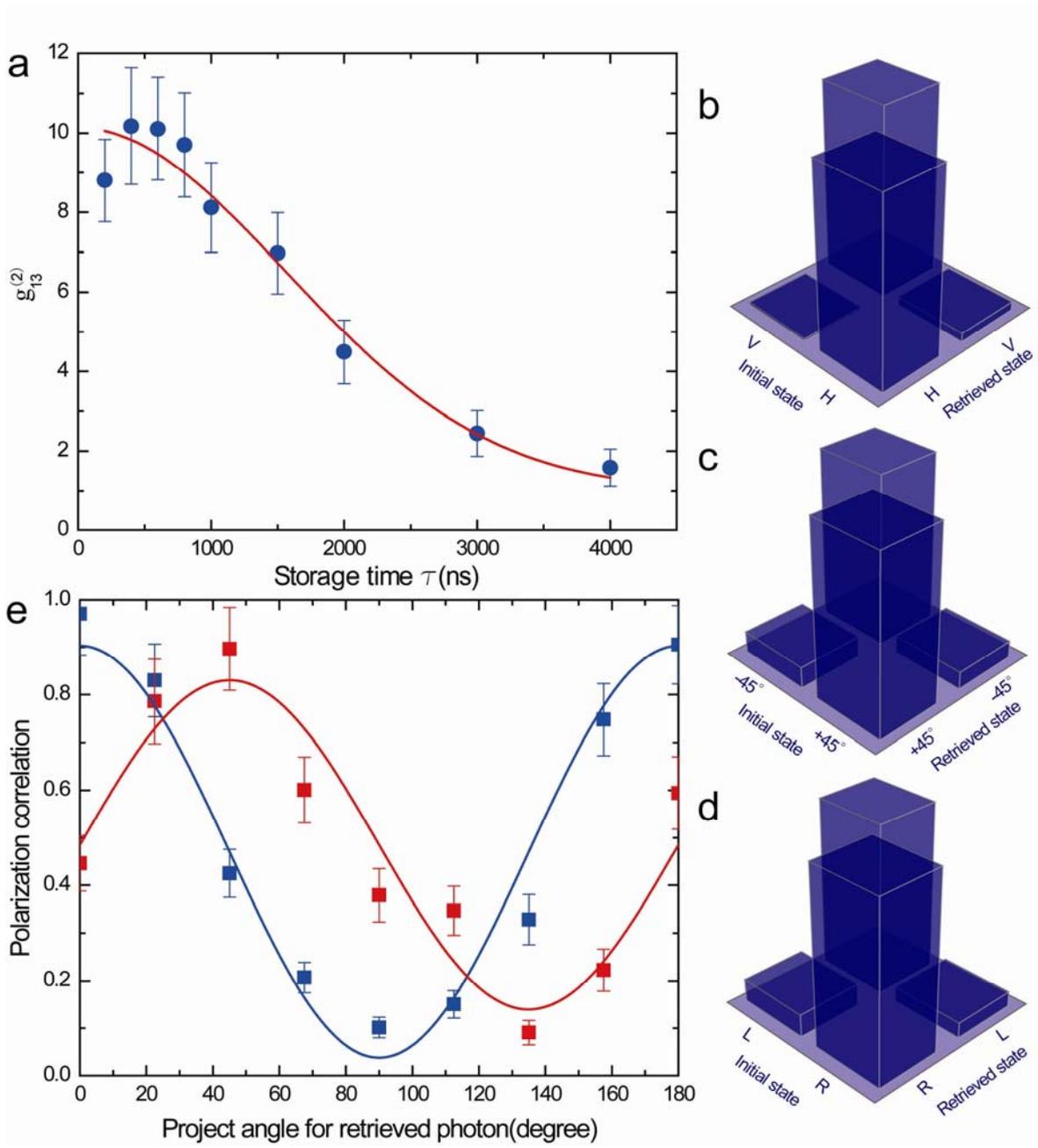

Figure-4

| Table 1 Measured violation of Bell inequality | | |
|---|---|---|
| Storage time(μs) | $S$ (CHSH) | Standard Deviation |
| 0(Local) | 2.59(5) | 11.8 |
| 0.2 | 2.54(13) | 4.2 |
| 1 | 2.28(17) | 1.7 |

Table-1